# Layer-dependent Band Alignment and Work Function of Few-Layer Phosphorene

Yongqing Cai, Gang Zhang[*] &Yong-Wei Zhang[§]

Institute of High Performance Computing, A*Star, Singapore 138632

*zhangg@ihpc.a-star.edu.sg; §zhangyw@ihpc.a-star.edu.sg

Using first-principles calculations, we study the electronic properties of few-layer phosphorene focusing on layer-dependent behavior of band gap, work function and band alignment and carrier effective mass. It is found that few-layer phosphorene shows a robust direct band gap character, and its band gap decreases with the number of layers following a power law. The work function decreases rapidly from monolayer (5.16 eV) to trilayer (4.56 eV), and then slowly upon further increasing the layer number. Compared to monolayer phosphorene, there is a drastic decrease of hole effective mass along the ridge (zigzag) direction for bilayer phosphorene, indicating a strong interlayer coupling and screening effect. Our study suggests that 1). Few-layer phosphorene with a layer-dependent band gap and a robust direct band gap character is promising for efficient solar energy harvest. 2). Few-layer phosphorene outperforms monolayer counterpart in terms of a lighter carrier effective mass, a higher carrier density and a weaker scattering due to enhanced screening. 3). The layer-dependent band edges and work functions of few-layer phosphorene allow for modification of Schottky barrier with enhanced carrier injection efficiency. It is expected that few-layer phosphorene will present abundant opportunities for a plethora of new electronic applications.

Recent years have witnessed remarkable progress in the application of two-dimensional (2D) materials in transistors and energy conversion[1,2]. Due to lack of band gap, which dramatically degrades the ON/OFF ratio, pristine graphene is incapable for high-performance field effect transistors. Consequentially, seeking other 2D materials that possess properties complementary to graphene is a fascinating subject of much active research. For example, studies have shown that 2D $MoS_2$ and some other transition metal dichalcogenide (TMD) family members are promising candidates due to their presence of a finite bandgap[3] and other intriguing electronic properties distinctive from graphene[4,5].

More recently, phosphorene, a 2D form of black phosphorus (BP), has attracted increasing attention due to its high carrier mobility, which is up to 1000 $cm^2V^{-1}s^{-1}$ at room temperature, and a moderate band gap, which is about 1.5 $eV$[6,7]. As the most stable phosphorus allotrope at ambient condition, bulk black phosphorus can be synthesized under 10 kbar pressure and high temperature, and phosphorene can be obtained by scotch tape-based mechanical exfoliation method[6]. In contrast to graphene and $MoS_2$, phosphorene is an intrinsic p-type semiconductor[8]. In addition, a substantial anisotropy in transport behavior was observed recently in few-layer phosphorene with thickness ranging from 2.1 to over 20nm[9]. Interestingly, recent studies showed that their anisotropic carrier mobility can be tuned by applying strain[10] and their electronic band gap can be engineered by the stacking order[11].

While a number of studies have been performed to understand the mechanical[12] and electronic properties of monolayer and few-layer phosphorene[13], some of their important properties, such as band alignment, work function and carrier effective mass, and their layer-dependent behavior, are still poorly understood. It is well-known that the energy difference between the valence (conduction) band edge of a semiconductor and the work function of a metal electrode determines the Schottky barrier at the metal-semiconductor interface. Hence, an in-depth understanding on the band alignment of the semiconductor is critical for high-performance electronic devices to achieve a high on-current and a low off-current[14]. However, the band alignment of few-layer phosphorene is still unclear. Moreover, constructing heterostructures consisting of individual monolayers in a precisely controlled stacking sequence with a precisely tuned band offset[15,16], is a highly promising route to fabricate 2D materials-based electronic devices. Since monolayer and few-layer phosphorene possess finite and yet moderate band gaps, they are of great potential for being used to construct these heterostructures with other 2D materials. Hence the knowledge on band alignment and work function of monolayer and few-layer phosphorene is indispensable. Unfortunately, both the band alignment and the work function and their thickness-dependent behavior are still unknown, which is clearly an obstacle to hinder their applications in nanoscale electronics and solar energy harvest[17].

In this work, we investigate the electronic properties of monolayer and few-layer phosphorene, focusing on their band alignment and work function. We find that a tunable work function up to 0.7eV can be achieved with increasing the thickness from monolayer (1L) to five layers(5L). In addition, there is a significant shift of valence band edge upon thickness variation, which can be useful for tuning Schottky barrier to promote hole injection efficiency. Our work suggests that by harnessing and engineering their thickness-dependent electronic properties, few-layer phosphorene can be promising for applications in high-performance devices.

## Results

**Thickness dependent band structure:** As shown in Fig. 1, the coordination number of each phosphorus atom is three, and the phosphorus atoms are covalently bonded to form a puckered 2D honeycomb structure. The puckered layer is composed of top and bottom ridges along the zigzag direction (**a**), and each two ridges are connected with surface-normal P-P bonds forming graphene-like armchair lattice (**b**). The relaxed lattice constant for the 1L phosphorene is a=3.34 Å and b=4.57 Å, in good agreement with previous study.[9] Similar to graphite, the multilayer BP is held together by van der Waals forces. Hereafter, for the bulk and few-layer BP, the structures are relaxed by using van der Waals corrected optB88 functional, which was shown to be the best estimation for the interlayer gap and binding energy in layered materials like graphene and BN among all the corrected functionals[18]. In the stacking direction, the ridges on the same side are directly atop each other, but there is a shift of **a**/2 between adjacent layers along the **x** direction. The cohesive energy $E_{coh}$ for the bilayer phosphorene is defined as $(E_{2L}-2E_{1L})/8$, where $E_{1L}$ and $E_{2L}$ are the total energy of the 1L and 2L phosphorene systems respectively, and 8 is the number of phosphorus atoms in the unit cell of 2L BP. The calculated cohesive energy $E_{coh}$ is -55 meV/atom, which is comparable to those of h-BN and graphite (both are around -65 meV/atom)[18].

To investigate the effect of different exchange-correlation functionals, we compare the band gap of single layer phosphorene using PBE, optB88, and HSE06 methods. The calculated values are 0.84 eV (PBE), 0.84 eV (optB88) and 1.52eV (HSE06). The band gap calculated using HSE06 is consistent with the experimental observation of 1.45 eV[9] and GW result of 1.6eV[13]. Owing to the inaccurate description of the electron-electron exchange interaction of the PBE and optB88 approaches, they predict a much smaller band gap (by about 0.7 eV). To better describe the exchange interaction, the band structures for bulk BP and few-layer BP are calculated by using HSE06 hybrid functional starting with the relaxed structure by optB88.

Figure 2 shows the band structure of bulk BP. A direct gap of 0.39 eV at Z point is revealed, which is in good agreement with previous experimental[19] and theoretical[20-23] works, thus validating the HSE06 functional in the description of band structure of BP. Figure 3 shows the band structures for 1L to 5L. Owing to the quantum confinement effect, the direct band gap at • point increases with reducing the thickness. For the bilayer case, the band gap is 1.02 eV, a large drop of 0.5 eV from the 1L case (1.52 eV). The topology of the band structure for other few-layer phosphorene is similar to that of the 1L case. For all the layers, there is a clearly asymmetric shape of the band dispersion around the • point along •-X and

•-Y branches, suggesting a large anisotropic behavior of electronic structures of few-layer phosphorene, which is in strong contrast to the isotropic electronic structures of graphene, BN and MoS$_2$ layers.

Fig. 4a shows the variations of the band gap with thickness from 1L to 5L. It is seen that there is a dramatic decrease in the band gap when the number of layer changes from 1L to 3L, whereas the value only decreases slightly when the number of layer is further increased. These results are in good agreement with previous studies[9,24]. By using the non-local correlation PBE and optB88 functional, we observe that although the values of calculated band gap are much smaller, the thickness dependence is consistent with that predicted by HSE06.

**A quantitative expression for layer-dependent band gap:** It is well-known that quantum confinement is able to increase the band gap of low-dimensional systems, such as nanowires and nanotubes. For example, it was shown that when the diameter of a nanowire decreases, its band gap increases by following[25-27], $E_g = E_0 + C/d^\alpha$, where $E_g$ is the band gap of the nanowire, $E_0$ is band gap for the bulk material, $d$ is the cylinder diameter of the nanowire, and $C$ and • are fitting parameters. Studies by using first principles calculations and semi-empirical methods showed that the parameter • is between 1 and 1.4 for one-dimensional systems[25-27]. To explore the corresponding quantum confinement effect in few-layer phosphorene, we examine the relationship between the band gap and the number of layers based on HSE06 calculation. The calculation results are shown in the inset of Fig. 4a in the log-log scale. It is seen clearly that there is a notable linear dependence, suggesting that the band gap of few-layer phosphorene also follows the same relation with the number of layers, that is:

$$E_g = E_0 + C/n^\beta \quad (1)$$

where $E_g$ is the band gap of few-layer phosphorene, $E_0$ is the band gap for bulk BP, which is calculated to be 0.39 eV, $n$ is the number of layers. The fitting parameters $C$ and • are 1.62 eV and 1.4, respectively. It is worth noting that this relationship is obtained based on the fitting of the band gap values of few-layer phosphorene from 2L to 5L. We find that monolayer phosphorene does not follow the above relationship. The underlying reason may be due to theabsence of interlayer coupling. Hence, this band gap expression is only applicable to few-layer phosphorene, in which the interlayer coupling plays in the critical role. It is known that the band gap of few-layer phosphorene is usually difficult to measure experimentally. Since thenumber of phosphorene layers can be obtained by optical measurement[15], the above expression provides a convenient formula to predict the band gap value of few-layer phosphorene.

**Thickness dependent work function and band alignment:** We calculate the work function by subtracting the Fermi energy from the electrostatic potential in the middle of vacuum[28, 29]. Since this procedure can be affected by the chosen exchange-correlation functional, it is expected that the predictions by using these functionals can be different. Fig. 4b shows the variation of the work function with the number of layers. It is seen that the work function for 1L phosphorene calculated by HSE06, PBE, and optB88 is 5.16, 4.50 and 4.43 eV respectively. Owing to the inaccuracy in considering the self-interaction of electrons by optB88 and PBE functionals, their predicted work functions are expected to be undervalued. Notably, there is a total reduction of the work function of 0.66, 0.43, and 0.57 eV for HSE06, PBE, and optB88, respectively, by increasing the number of layers from 1L to 5L. Based on the predictions by HSE06 approach, we see that the work function decreases rapidly from 1L(5.16 eV) to 3L (4.56 eV)by 0.6 eV, and then slowly from 3L (4.56 eV) to 5L (4.5 eV) by only 0.06 eV, showing the same trend as the band gap change as shown in Fig.4a. This remarkable change in the work function with the number of layers suggests the possibility to tune the contact resistance by controlling the number of phosphorene layers and selecting suitablecontact metals.

Figure 5 shows the alignment of valence band maximum (VBM) and conduction band minimum (CBM) with respect to the vacuum level for 1L to 5L based on the HSE06 calculation. It can be seen that the position of the VBM for 1L is located at -5.43eV, and for 3L at -4.85 eV, and for 5L at -4.70 eV; and the position of the CBM for 1L is located at -3.91 eV, and for 3L at-4.09 eV and for 5L at -4.13 eV. These results clearly show that there is a rapid upward shift of the VBM and a slow downward shift of CBM. Hence, the adjustment of the VBM with the number of layers is much more intensive than that of the CBM. As an intrinsic p-type semiconductor, the strong layer number-dependent position of VBM in few-layer phosphorene offers a practical route to tune the Schottky barrier height, which can lead to more efficient hole injection and hole transport across the contact.

## Discussion

Doping has been widely used to control the electronic properties of semiconductors. A field effect transistor (FET) can be either n- or p-type, depending on whether electrons or holes are the majority carriers contributed by the dopants. Meanwhile, dopant can distort the lattice of a semiconductor, resulting in enhanced scattering and thus reduced carrier mobility. In recent years, a doping-free technique has been developed in nanoscale electronics, which allows carriers to be supplied directly from the metal electrodes[30]. For example, when a metal with a high work function is used as electrode, a p-type carbon nanotube FET can be achieved via the Ohmic contact between carbon nanotube valence band and metal[30]. Conversely, when a metal with a low work function is used as electrode, a n-type carbon nanotube FET can be obtained via the Ohmic contact between the carbon nanotube conduction band and metal. Thus it is the relative position between the conduction/valence band edge of semiconductor and the Fermi level of metal electrode that determines the fundamental characteristics of these FETs. In Fig.5, for comparison, we also plot the work function of a few metals like Ti, Cu, Ni and Pt, which have been widely used as the electrodes. Remarkably, the tunable conduction/valence band edge of few-layer phosphorene with the number of layers potentially provides a practical means to enhance the performance of its contact with various metal electrodes. However, for real applications, the environmental effects due to impurity and substrate may alter the electronic properties and the work functions of few-layer phosphorene. In addition, as a layered nanomaterial, few-layer BP may suffer from uncertainties arising from the stacking disorder induced by the relative displacement and twisting between the layers. For example, previous study has shown that the stacking variation in bilayer phosphorene can greatly modify the band gap while the direct gap character seems to be unaffected[11].

Fig. 3 and Fig. 5 clearly show the layer number-dependent bandgap of few-layer phosphorene (from 1.52 to 0.57 eV by HSE06 method). Hence, few-layer phsophorene can be potentially used to constructmulti-junction solar cells, which can significantly enhance the energy conversion efficiency by covering the full solar spectrum[31]. Recent experiment demonstrated that BP transistors indeed respond fast to excitation wavelength ranging from the visible region up to 940 nm[32]. Importantly, monolayer and few-layer phosphorene all exhibit a direct bandgap character, which are in contrast to monolayer and few-layer TMD materials, which usually exhibit a thickness-induced direct-to-indirect transition[3]. This robust electronic character of monolayer and few-layer phosphorene is ideally suitable for those applications which demand both a high carrier concentration and a good light adsorption efficiencysimultaneously.

Finally, we examine the thickness-dependent behavior of the effective mass by using the formula $\hbar^2[\partial^2\varepsilon(k)/\partial k_\alpha \partial k_\beta]^{-1}$. It should be noted that the calculated effective masses from the HSE06 method is not plotted since it predicts an oscillating effective mass trend with the number of layers due to the substitution of part of the semi-local exchange-correlation functional with the Hartree-Fock exchange. Hence, here, the mass tensor is derived from the calculations by the PBE method. The obtained electron/hole masses along **a** and **b** directions are shown in Fig. 6. It is seen that there is a strong anisotropy between hole and electron masses along **a** and **b** directions. The electron mass along **a** (•-X) direction is significantlylarger (by one order) than that along **b** (•-Y) direction for all the calculated layers. The electron mass of 1L is 1.23 $m_e$ along the •-X direction and 0.14 $m_e$ along the •-Y direction, in agreement with previous study[10]. A similar behavior is also observed for holes. For 1L, the anisotropy of hole mass is extraordinarily large: a two-order difference between the two directions. This strong orientation-dependent effective masses of electron/hole of 1L phosphorene provides a plausible explanation for the large anisotropic conducting behavior reported recently[9]. Such strong in-plane anisotropic behavior, which is not found in other 2D materials like BN or $MoS_2$, suggests possible novel applications for phosphorene. It is interesting to note from Fig. 6a that there is a dramatic decrease of hole mass in 1L phosphorene (13.09 $m_e$) to 2L phosphorene (2.18 $m_e$) along **a** direction, suggesting a significant change of the band curvature due to the strong interlayer coupling. Since a decrease in the carrier mass generally implies a higher mobility, few-layer phosphorene is likely to perform better than monolayer phosphorene.

In summary, we have investigated the electronic properties of few-layer phosphorene by first-principles calculations. It is found that the band gap, band alignment, work function and carrier effective masses are all dependent on the number of layers. A simple expression is proposed to predict the variation of band gap with the number of phosphorene layers. Importantly, the varied band gap and robust direct band gap character in few-layer phosphorene provides a basis to fabricate multi-junction solar cells which can more effectively harvest solar energy. The layer number-dependent band alignment and work function offer a practical route to tune the Schottky barrier height, which can lead to more efficient hole injection and hole transport across the contact. There is a clear decreasing trend for carrier effective masses with increasing the number of layers, suggesting a higher mobility for multilayer phosphorene than monolayer phosphorene. This is especially true for the hole mass which exhibits a significant thickness dependence along the ridge direction due to the interlayer coupling and screening. In addition, few-layer phosphorene can have an enhanced carrier density and an improved structural stability due to the enhanced interlayer coupling and screening, it is expected that few-layer phosphorene will present abundant opportunities for a plethora of new electronic applications.

## Methods

The atomic structure and the stacking order of few-layer phosphorene are shown in Fig. 1.In the present work, the number of few-layer phosphorene is denoted by $n$L, where $n$ is the number of the monolayer phosphorene, and we consider 1L to 5L (0.66 nm to 2.8 nm). All our calculations are performed using the planewave code Vienna ab initio simulation package (VASP)[33] within the framework of density functional theory (DFT). Spin-restricted generalized-gradient approximation with the Perdew-Burke-Ernzerhof functional (PBE)[34] and hybrid functionals (HSE06)[35] are utilized, together with the projector augmented wavemethod. The thickness of the vacuum region is greater than 15Å. The first Brillouin zone is sampled with a 12×10×1 Monkhorst-Pack grid and a kinetic energy cutoff of 500 eV is adopted. The PBE approach is well-known to have an inaccurate description of the dispersion force and thus a poor estimation of the interlayer distance. Here all the structures are relaxed by using the optimized Becke88 van der Waals (optB88-vdW) functional[36]. The lattice constant is relaxed until the stress exerted on the cell is less than 0.1 kbar and the forces exerted on the atoms are less than 0.01 eV/Å.

## Acknowledgments

The authors gratefully acknowledge the financial support from the Agency for Science, Technology and Research (A*STAR), Singapore and the use of computing resources at the A*STAR Computational Resource Centre, Singapore.

## Author contributions

Y. C. performed the first-principles calculations. G. Z. and Y.-W. Z. contributed to the discussion and writing of the manuscript. All authors reviewed the manuscript.

## Additional information

Competing financial interests: The authors declare no competing financial interests.

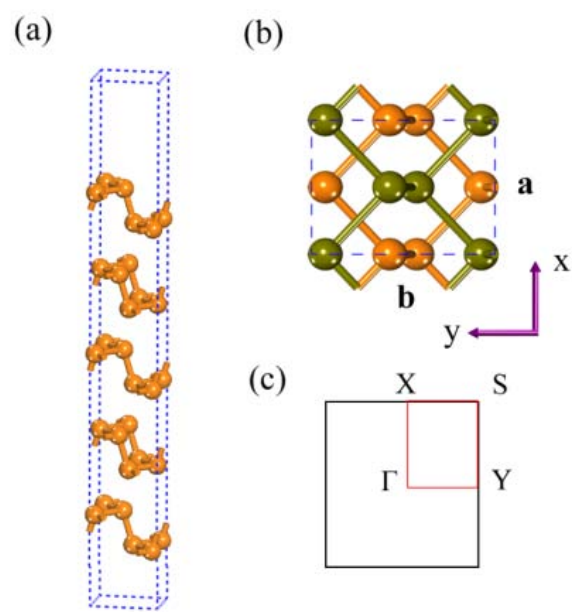

Figure 1. (a) Side view of the atomic structure of 5L phosphorene. (b) Top view of bilayer phosphorene, where the atoms in different color representing atoms in different layers. (c) The 2D Brillouin zone for few-layer phosphorene.

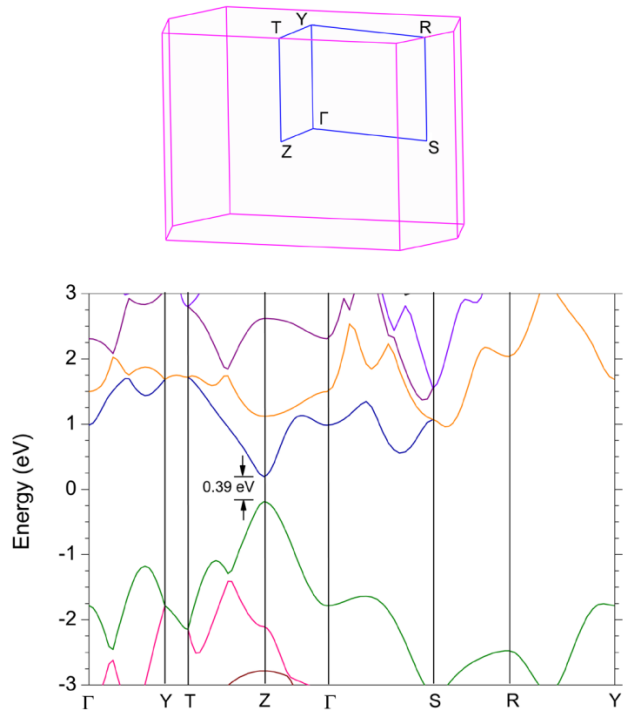

Figure 2. Band structure of the bulk BP calculated by the HSE06 method. The 2D Brillouin zone is shown in the inset.

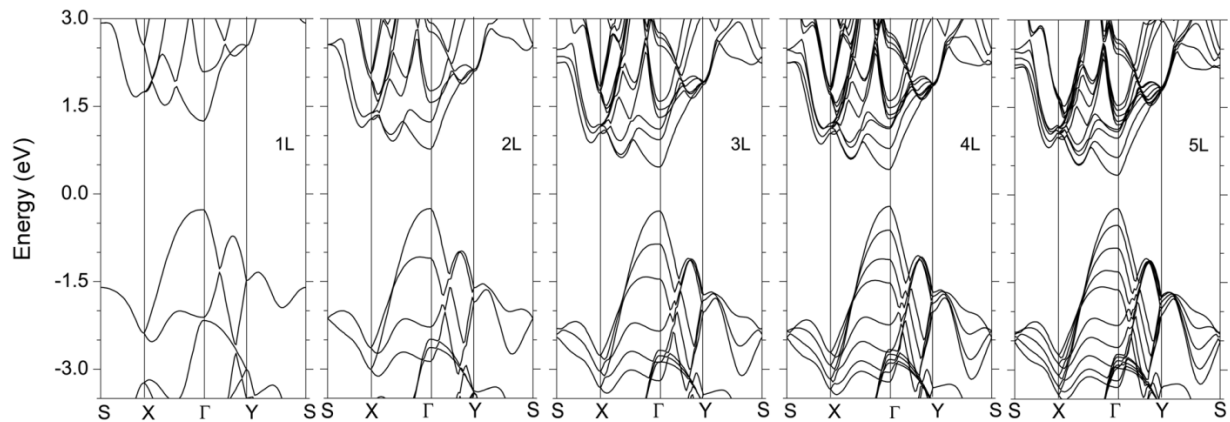

Figure 3. Band structures for different few-layer phosphorene systems obtained from HSE06 hybrid functional calculations. Note that the direct band gap character is maintained for all thicknesses up to 5L, in contrast to $MoS_2$, which is of a direct band gap character only for monolayer[3].

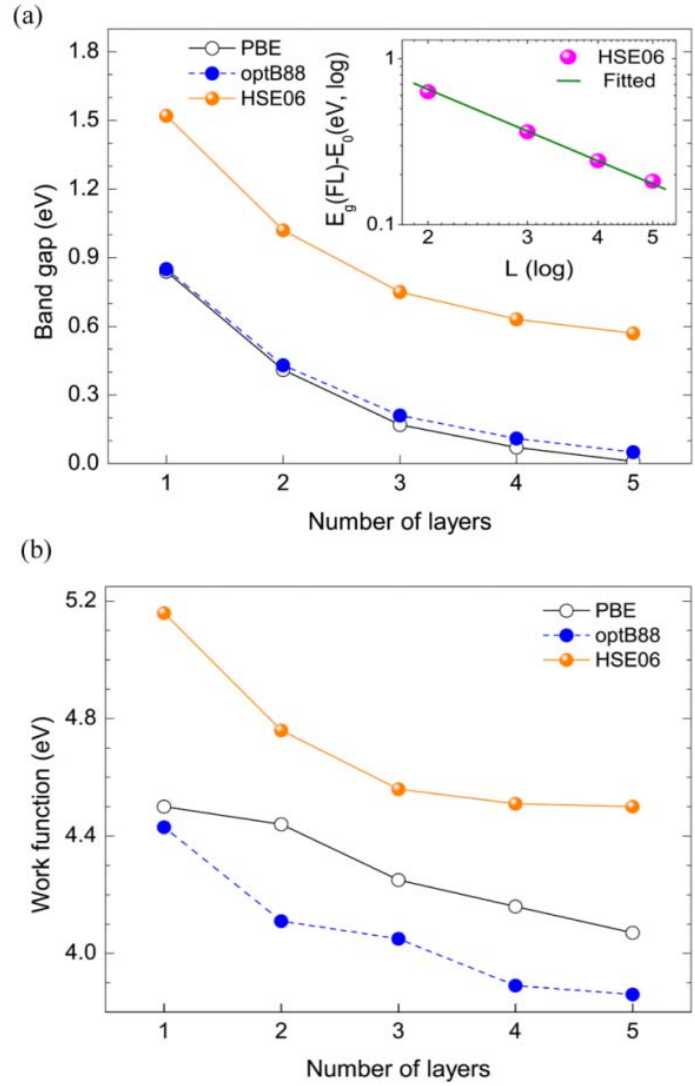

Figure 4. Variation of the band gap (a) and work function (b) with the thickness of few-layer phosphorene. The inset of (a) shows the band gap versus the number of layers in the log-log scale.

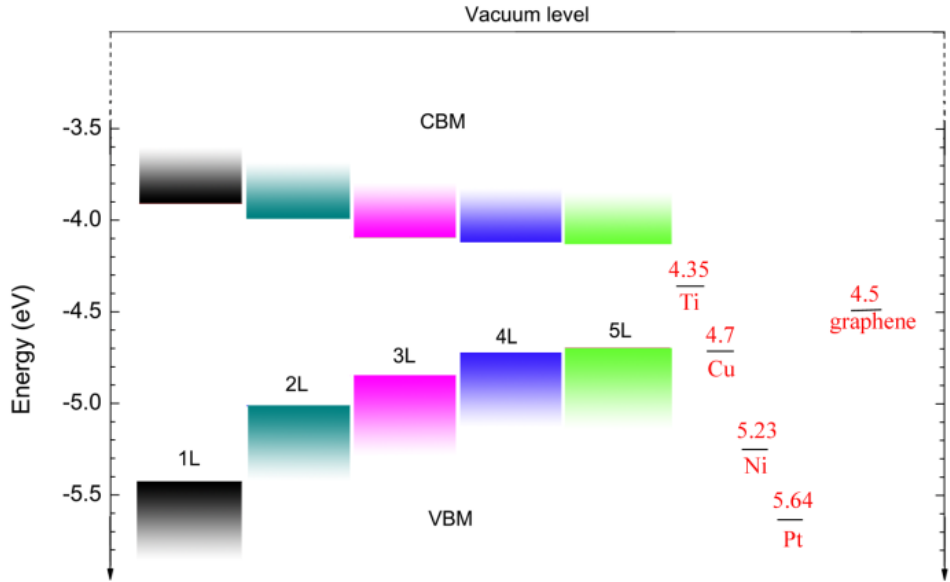

Figure 5. Variation of valence band maximum (VBM) and conduction band minimum (CBM) for few-layer phosphorene with the number of layers (from 1L to 5L) determined from HSE06 calculation. The VBM and CBM are calculated with respect to the vacuum level. The value of the work function for graphene is taken from Ref. 1 and that for metals from Ref. 29.

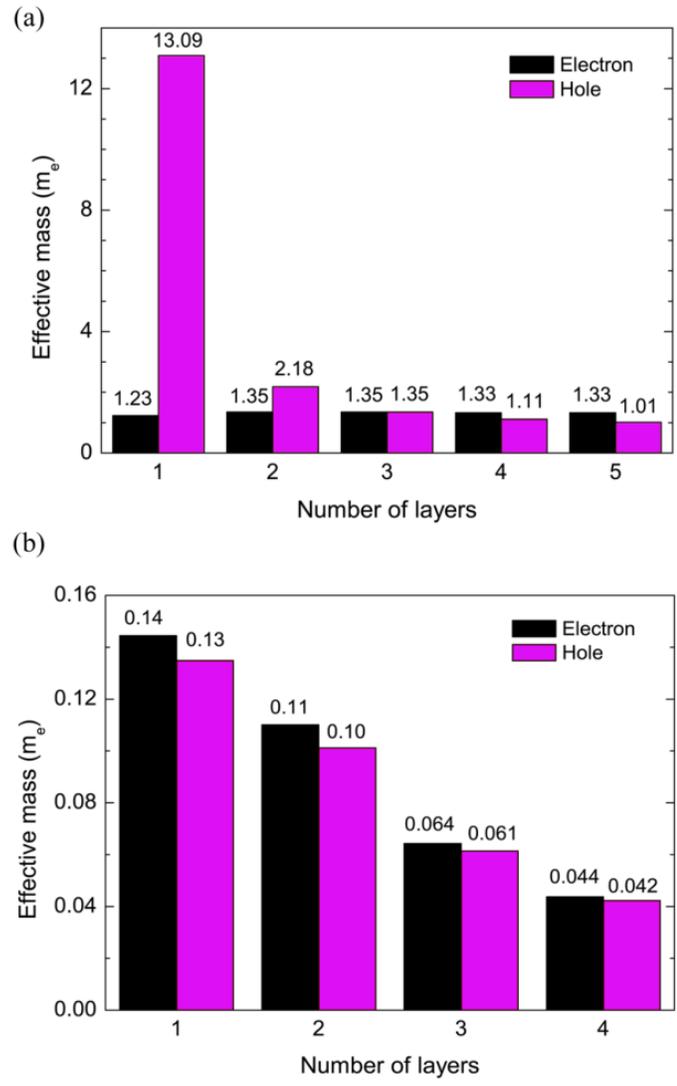

Figure 6.Thickness-dependent effective masses for holes and electrons along **a** (a) and **b** (b) direction in few-layer phosphorene.